\documentclass[11pt,a4paper]{article}

\usepackage{jheppub}
\usepackage{amsfonts}
\usepackage{amssymb}
\usepackage{comment}
\usepackage{epsfig}
\usepackage{graphicx}
\usepackage{amsmath}
\usepackage{tabu}
\usepackage{subfig}
\usepackage{float}

\newcommand{\bea}{\begin{eqnarray}}
\newcommand{\eea}{\end{eqnarray}}
\newcommand{\bi}{\begin{itemize}}
\newcommand{\ei}{\end{itemize}}
\newcommand{\ben}{\begin{enumerate}}
\newcommand{\een}{\end{enumerate}}
\newcommand{\be}{\begin{equation}}
\newcommand{\ee}{\end{equation}}
\newcommand{\ba}{\begin{align}}
\newcommand{\ea}{\end{align}}
\newcommand{\comments}[1]{}

\newcommand\vo{{\mathcal{V}}}
\newcommand\mc{\mathcal}
\newcommand{\beqa}{\begin{eqnarray}}
\newcommand{\eeqa}{\end{eqnarray}}
\newcommand{\V}{{\cal{V}}}
\title{Microscopic Origin of Volume Modulus Inflation}

\author[1,2,3]{Michele Cicoli,}
\author[2,3]{Francesco Muia,}
\author[4,5]{Francisco Gil Pedro}

\affiliation[1]{ICTP, Strada Costiera 11, Trieste 34014, Italy}
\affiliation[2]{Dipartimento di Fisica e Astronomia, Universit\`a di Bologna, \\ via Irnerio 46, 40126 Bologna, Italy}
\affiliation[3]{INFN, Sezione di Bologna, via Irnerio 46, 40126 Bologna, Italy}
\affiliation[4]{Departamento de Fisica Te\'orica UAM and Instituto de Fisica Te\'orica UAM/CSIC}
\affiliation[5]{Universidad Aut\'onoma de Madrid, Cantoblanco, 28049 Madrid, Spain}

\emailAdd{mcicoli@ictp.it}
\emailAdd{muia@bo.infn.it}
\emailAdd{francisco.pedro@csic.es}

\abstract{High-scale string inflationary models are in well-known tension with low-energy supersymmetry. A promising solution involves models where the inflaton is the volume of the extra dimensions so that the gravitino mass relaxes from large values during inflation to smaller values today. We describe a possible microscopic origin of the scalar potential of volume modulus inflation by exploiting non-perturbative effects, string loop and higher derivative perturbative corrections to the supergravity effective action together with contributions from anti-branes and charged hidden matter fields. We also analyse the relation between the size of the flux superpotential and the position of the late-time minimum and the inflection point around which inflation takes place. We perform a detailed study of the inflationary dynamics for a single modulus and a two moduli case where we also analyse the sensitivity of the cosmological observables on the choice of initial conditions.}


\preprint{IFT-UAM/CSIC-15-100}

\begin{document}

\maketitle

\section{Introduction}

Two crucial quantities of any string compactification are the Hubble scale during inflation $H$ and the gravitino mass $m_{3/2}$. The first sets the inflationary energy scale $M_{\rm inf}\sim \sqrt{H M_P}$ which in turn is related to the tensor-to-scalar ratio $r$ as $M_{\rm inf}\sim M_{\rm GUT}\left(\frac{r}{0.1}\right)^{1/4}$, whereas the second gives the order of magnitude of the soft supersymmetry breaking terms $M_{\rm soft}\sim m_{3/2}$. Given that neither primordial tensor modes nor supersymmetric particles have been detected yet, experimental data yield just upper and lower bounds on these two quantities: 
\bea
r&\lesssim& 0.1 \qquad\quad\,\Rightarrow \qquad H \lesssim \frac{M_{\rm GUT}^2}{M_P} \sim 10^{14}\,{\rm GeV} \label{rbound} \\
M_{\rm soft}&\gtrsim& 1\,{\rm TeV}\qquad\Rightarrow\qquad m_{3/2}\gtrsim 1\,{\rm TeV}\,. \label{Msoftbound}
\eea
However $H$ and $m_{3/2}$ are not two independent quantities since in any consistent string inflationary model the inflaton dynamics has not to destabilise the volume mode. This is guaranteed if the inflationary energy is smaller than the energy barrier to decompactification, i.e. $H^2 M_P^2 \lesssim V_{\rm barrier}$, and the height of the barrier is generically set by the gravitino mass. In KKLT models $V_{\rm barrier}\sim m_{3/2}^2 M_P^2$ which leads to $H\lesssim m_{3/2}$ \cite{Kallosh:2004yh} while in the LARGE Volume Scenario (LVS) $V_{\rm barrier}\sim m_{3/2}^3 M_P$ giving $H\lesssim m_{3/2}\sqrt{\frac{m_{3/2}}{M_P}}$ \cite{Conlon:2008cj}.

These theoretical bounds are not in contradiction with the experimental bounds (\ref{rbound}) and (\ref{Msoftbound}), in particular for the cases of high-scale supersymmetry and small field inflationary models with unobservable tensor modes. However the phenomenologically interesting cases of low-energy supersymmetry with $M_{\rm soft}\gtrsim \mc{O}(1)$ TeV and large field inflationary models with $r\gtrsim \mc{O}(0.01)$ would imply a high value of $H$ together with a small value of $m_{3/2}$, in clear tension with the theoretical bounds from volume destabilisation problems.

Several ideas have been proposed in the literature to overcome this tension between TeV-scale supersymmetry and large field inflation. Here we briefly summarise them:
\ben
\item The relation $H^2 M_P^2\lesssim V_{\rm barrier}$ would be independent on $m_{3/2}$ in models where the energy barrier and the gravitino mass are two uncorrelated quantities. Racetrack superpotentials provide viable models where $V_{\rm barrier}$ is decoupled from $m_{3/2}$ \cite{Kallosh:2004yh}.

\item The KKLT bound $H\lesssim m_{3/2}$ and the LVS bound $H\lesssim m_{3/2}\sqrt{\frac{m_{3/2}}{M_P}}$ apply just to the inflationary era, and so in these expression $m_{3/2}$ is the gravitino mass during inflation which might be different from the present value of the gravitino mass. This is possible if $m_{3/2}= e^{K/2} |W| \simeq \frac{W_0}{\vo} M_P$ evolves just after the end of inflation. Two viable realisations include inflationary models where the inflaton coincides with the volume mode $\vo$ so that $\vo$ relaxes from small to large values during inflation \cite{Conlon:2008cj} or where the inflaton is a complex structure modulus or a matter field so that $W$ relaxes from large to small values during inflation \cite{He:2010uk}. Another option is to consider models with two different volume stabilisation mechanisms during and after inflation. If $\vo$ couples to the field $X$ whose F-term drives inflation, the volume's vacuum expectation value (VEV) during inflation would be determined by the F-term potential of $X$ which however vanishes after the end of inflation when $\vo$ is fixed by a more standard KKLT or LVS mechanism. This kind of models with a K\"ahler potential coupling between $\vo$ and $X$ have been studied in \cite{Antusch:2011wu} whereas superpotential interactions between $\vo$ and $X$ have been analysed in \cite{Yamada:2012tj}.

\item Another possible way-out to reconcile low-energy supersymmetry with high scale inflation is to consider models where the visible sector is sequestered from the sources of supersymmetry breaking so that the soft terms are much smaller than the gravitino mass \cite{Seq}. This can be the case for compactifications where the visible sector is localised on fractional D3-branes at singularities which can lead to $M_{\rm soft}\sim m_{3/2}\frac{m_{3/2}}{M_P}$.
\een 

Let us point out that all the solutions listed above require a high degree of tuning except for the sequestered case which might be however not enough to completely remove all the tension between observable tensor modes and TeV-scale supersymmetry. Moreover, it is technically rather complicated to provide consistent string models where these solutions are explicitly realised. Therefore they are at the moment at the level of string-inspired toy-models without a concrete string embedding where one can check if there is enough freedom to achieve the amount of fine tuning needed to reproduce all the desired phenomenological details.

In this paper we shall provide a first step towards an explicit stringy embedding of the case where the volume mode plays the r\^ole of the inflaton evolving from small to large values after the end of inflation \cite{Conlon:2008cj}. We will describe a possible microscopic origin of the potential terms used in \cite{Conlon:2008cj} to create an inflection point at small volumes around which slow-roll inflation can occur. Moreover, we shall also perform a deeper analysis of the relation between the positions of the inflection point and the minimum with the tuning of the flux superpotential.

Before presenting the details of our analysis, let us stress some key-features of the model under study:
\bi
\item In order to have an evolving gravitino mass, we focus on models where the inflaton is the volume mode $\vo$. Given that we work in an effective supergravity theory where the K\"ahler potential has a logarithmic dependence on $\vo$, each term in the inflationary potential will be a negative exponential of the canonically normalised volume mode $\Phi\sim \ln\vo$, i.e. $V \supset e^{- k \Phi}$. This form of the potential is reminiscent of Starobinsky-like models which have a rather large inflationary scale since they are at the boarder between large and small field models \cite{Starobinsky:1980te}. However Starobinsky-like potentials feature also an inflaton-independent constant which can never appear in cases where the inflaton is the volume mode $\vo$ since $\vo$ couples to all sources of energy because of the Weyl rescaling needed to obtain the correct effective action.\footnote{Starobinsky-like models with large $r$ and a constant inflaton-independent constant can instead be obtained in models where the inflaton is a K\"ahler modulus different from the volume mode \cite{Cicoli:2008gp}.}

\item The best way to achieve volume inflation is therefore to consider a potential which has enough tuning freedom to create an inflection point and then realise inflation in the vicinity of the inflection point. The price to pay is that this inflationary scenario turns out to be rather fine tuned and it is necessarily a small field model with a sub-Planckian field range during inflation $\Delta\Phi\sim 0.4 \,M_P$, unobservable tensor modes of order $r\sim 10^{-9}$ and low Hubble scale $H\sim 10^{10}$ GeV. Hence this approach cannot solve completely the tension between large tensor modes and TeV-scale supersymmetry. Still it provides a big step forward especially if combined with a sequestered visible sector so that low-energy supersymmetry can be safely reconciled with a high gravitino mass.

\item In LVS models where $H> m_{3/2} \sqrt{\frac{m_{3/2}}{M_P}}$ and the inflaton is the volume mode, the destabilisation problem of \cite{Kallosh:2004yh} becomes an overshooting problem since the inflaton has an initial energy which is larger than the barrier to decompactification. The solution to this problem via radiation production after the end of inflation has already been discussed in \cite{Conlon:2008cj}, and so we shall not dwell on this issue. 

\item In the models under study, the Hubble scale during inflation is set by the gravitino mass during inflation $m_{3/2}^{\rm inf}$ which is much larger than the gravitino mass today $m_{3/2}^{\rm today}$ due to the volume evolution. Hence the Hubble scale $H$ can be much larger than $m_{3/2}^{\rm today}$ since we have:
\be
H \sim m_{3/2}^{\rm inf} \sqrt{\frac{m_{3/2}^{\rm inf}}{W_0\,M_P}} \gg m_{3/2}^{\rm today} \sqrt{\frac{m_{3/2}^{\rm today}}{W_0\,M_P}}\,.
\ee
\ei
We shall analyse both the single modulus and the two moduli case focusing on three different visible sector realisations which lead to:
\ben
\item \textit{High-scale SUSY models}: in this case the requirement of low-energy supersymmetry is abandoned and the value of the gravitino mass both during and after inflation is huge. The volume mode evolves from values of order $100$ during inflation to values of order $200$ after the end of inflation. The flux superpotential has to be tuned to values of order $W_0\sim 10^{-5}$ in order to reproduce the correct amplitude of the density perturbations. Thus the order of magnitude of the gravitino mass during and after inflation is the same, $m_{3/2}^{\rm inf}\sim m_{3/2}^{\rm today}\sim 10^{11}$ GeV corresponding to $H\sim 10^{10}$ GeV. Due to the small value of $\vo$, in this case the validity of the effective field theory approach is not fully under control.

\item \textit{Non-sequestered models}: in these models during inflation the volume is of order $\vo \sim 10^5$ and $W_0\sim 1$ giving a gravitino mass of order $m_{3/2}^{\rm inf}\sim 10^{14}$ GeV which leads again to $H\sim 10^{10}$ GeV. After the end of inflation the volume evolves to $\vo\sim 10^{15}$ as required to get TeV-scale supersymmetry since the present value of the gravitino mass becomes $m_{3/2}^{\rm today}\sim 10$ TeV \cite{Conlon:2005ki}.

\item \textit{Sequestered models}: in these models inflection point volume inflation takes place again for values of order $\vo \sim 10^5$ and $W_0\sim 1$ which yield $m_{3/2}^{\rm inf}\sim 10^{14}$ GeV and $H\sim 10^{10}$ GeV. After the end of inflation the volume evolves instead to $\vo\sim 10^7$ corresponding to $m_{3/2}^{\rm today}\sim 10^{11}$ GeV, as required to get low-energy supersymmetry gaugino masses in sequestered scenarios where $M_{1/2}\sim m_{3/2} \frac{m_{3/2}}{M_P}\sim 10$ TeV \cite{Seq}. 
\een

This paper is organised as follows. Sec. \ref{SecIPInf} is a brief review of the basic concepts of inflection point inflation which will be used in the rest of the paper while in Sec. \ref{SecOrigin} we describe a possible microscopic origin of all the terms in the inflationary potential which are needed to develop an inflection point at small values of $\vo$ together with a dS minimum at larger values of the volume mode. In Sec. \ref{SecSingle} we study the single modulus case presenting first an analytical qualitative description of the inflationary dynamics and then performing an exact numerical analysis. The two moduli case typical of LVS models is instead discussed in Sec. \ref{SecTwo} before presenting our conclusions in Sec. \ref{SecConcl}.

\section{Inflection point inflation}
\label{SecIPInf}

In this section we briefly review the generic features of inflection point inflation, closely following  \cite{Baumann:2007ah} and \cite{Linde:2007jn}. We summarise the main points and discuss the tuning involved in these models.

The basic assumption is that inflation takes place around an inflection point along some arbitrary direction in field space. The scalar potential around such an inflection point can always be expanded as:
\be
V = V_{\rm ip} \left(1+ \lambda_1 (\phi - \phi_{\rm ip})+ \frac{\lambda_3}{3!} (\phi - \phi_{\rm ip})^3+\frac{\lambda_4}{4!} (\phi - \phi_{\rm ip})^4+\dots\right),
\ee
where $\phi_{\rm ip}$ denotes the position of the inflection point. Given that we shall focus on cases where the field excursion during the inflationary period is small, i.e. $(\phi - \phi_{\rm ip})\ll 1$, the quartic term can be safely neglected. We therefore find that it suffices to analyse the following potential:
\be
V = V_{\rm ip} \left(1+ \lambda_1 (\phi - \phi_{\rm ip}) + \frac{\lambda_3}{3!} (\phi - \phi_{\rm ip})^3\right).
\label{appr}
\ee
The inflationary observables are determined by the slow-roll parameters:
\bea
\epsilon &=& \frac{1}{2} \left(\frac{V'}{V}\right)^2 = \frac{1}{2} \left(\lambda_1 + \frac{1}{2} \lambda_3 (\phi - \phi_{\rm ip})^2\right)^2\simeq \frac{1}{2}\lambda_1^2\,, 
\label{eps} \\
\eta &=& \frac{V''}{V} = \lambda_3 (\phi - \phi_{\rm ip})\,,
\label{eq:srParams}
\eea
which have to be evaluated at horizon exit where $\phi=\phi_*$. Note that in (\ref{eps}) and (\ref{eq:srParams}) we approximated $V \simeq V_{\rm ip}$ in the denominator, which is a good approximation for small field models where $(\phi - \phi_{\rm ip})\ll 1$. Notice that if $\lambda_3\gtrsim 1$ this small field condition has to be satisfied in order to obtain $\eta\ll 1$. If instead $\lambda_3$ is tuned such that $\lambda_3\ll 1$, the condition $\eta\ll 1$ could be satisfied also for large field values but then the approximation (\ref{appr}) would be under control only by tuning all the coefficients of the expansion. We shall therefore focus only on the case $(\phi - \phi_{\rm ip})\ll 1$. The number of e-foldings is given by:
\be 
N_e(\phi_*) = \int_{\phi_{\rm end}}^{\phi} \frac{d \phi}{\sqrt{2 \epsilon}} = \sqrt{\frac{2}{\lambda_1 \lambda_3}} 
\arctan \left.\left[\sqrt{\frac{\lambda_3}{2\lambda_1}}(\phi - \phi_{\rm ip})\right]\right|_{\phi_{\rm end}}^{\phi_*}\,.
\label{Ne}
\ee
In order to have enough e-foldings we need $\lambda_1\ll 1$, which is also needed to get $\epsilon\ll 1$, and $(\phi - \phi_{\rm ip})\gtrsim \sqrt{\lambda_1}$ so that the arctangent does not give a small number. Thus the slow-roll parameter $\eta$ turns out to be larger than $\epsilon$ since:
\be
\epsilon\sim \lambda_1^2 \ll \eta \sim (\phi - \phi_{\rm ip})\gtrsim \sqrt{\lambda_1}\ll 1\,.
\ee
Therefore the spectral index in these models is essentially given by $\eta$:
\be
n_s - 1 = 2 \eta(\phi_*) - 6 \epsilon(\phi_*) \simeq 2 \eta(\phi_*)\,,
\ee
By using (\ref{Ne}), it is possible to rewrite the spectral index as a function of the number of e-foldings as: 
\be
n_s - 1 \simeq - \frac{4}{N_e} + \frac 23 \lambda_1\lambda_3 N_e\,.
\label{eq:ns}
\ee
Since $1 - \frac{4}{N_e} \simeq 0.93$ for $N_e\simeq 60$, it is evident that for a very small $\lambda_1$ (or equivalently a very flat inflection point) the spectral index asymptotes to $0.93$. We have therefore to use (\ref{eq:ns}) to determine the value of $\lambda_1$ that gives a value of $n_s$ in agreement with recent Planck data, i.e. $n_s = 0.9655\pm 0.0062$ ($68\%$ CL) \cite{Ade:2015lrj}. We find:
\be
n_s=0.965 \quad\Rightarrow\quad \lambda_1=7.92\cdot 10^{-4}\,\lambda_3^{-1}\qquad\text{for}\qquad N_e=60\,.
\label{correctns}
\ee
In these small field models, the tensor-to-scalar ratio $r$ turns out to be unobservable since:
\be
r= 16 \epsilon \simeq 8\lambda_1^2 \simeq 5.01\cdot 10^{-6} \lambda_3^{-2}\qquad\text{for}\qquad N_e=60.
\label{r}
\ee
Successful inflation not only gives rise to the correct spectral tilt and tensor fraction but does so at the right energy scale. The normalisation of scalar density perturbations in these models can be written as: 
\be
\Delta^2 = \frac{1}{24 \pi^2} \left.\frac{V}{\epsilon}\right|_{\phi_*} \simeq \frac{1}{12 \pi^2} \frac{V_{\rm ip}}{\lambda_1^2} \simeq 2.4 \cdot 10^{-9}\,.
\label{cobe}
\ee
Once the parameter $\lambda_3$ is fixed, (\ref{correctns}) gives the value of $\lambda_1$ which produces the correct spectral index, $n_s=0.965$, and (\ref{cobe}) fixes the value of $V_{\rm ip}$ which reproduces the observed amplitude of the density perturbations. In turn, (\ref{r}) yields the prediction for the tensor-to-scalar ratio $r$.

\section{Microscopic origin of the inflationary potential}
\label{SecOrigin}

In this section we will try to describe a possible microscopic origin of volume modulus inflation focusing on tree-level effects first, and then including any subleading effect which breaks the no-scale structure typical of type IIB constructions. 

\subsection{Tree-level effective potential}
\label{sec:pot}

The closed string moduli content of the low-energy supergravity limit of type IIB string theory compactified on a Calabi-Yau $X$ can be summarised as:
\begin{itemize}
\item Dilaton $S = s + {\rm i} C_0$
\item Complex structure moduli $U_\alpha$ ($\alpha = 1, \dots, h^{1,2}(X)$)
\item K\"ahler moduli $T_i = \tau_i + {\rm i} c_i$ ($i = 1, \dots, h^{1,1}(X)$)
\end{itemize}
where $C_0$ and $c_i$ are axion fields, while $U_\alpha$ are complex functions. It is worth recalling that the VEV of the real part of the dilaton sets the string coupling: $\langle s\rangle = 1/g_s$.

The effective field theory is completely determined by the K\"ahler potential, the superpotential and the gauge kinetic functions. While the gauge kinetic functions are model-dependent, a generic feature of type IIB compactifications is the form of the tree-level K\"ahler potential:
\be
K_{\rm tree} = - 2 \ln\vo - \ln\left(S + \overline{S}\right) - \ln\left(i \int_X \Omega \wedge \overline{\Omega}\right).
\label{leadingKP}
\ee
Turning on background fluxes induces the Gukow-Vafa-Witten superpotential \cite{Gukov:1999ya}:
\be
W_0(U,S) = \int_X G_3 \wedge \Omega\,,
\ee
where $G_3 = F_3 + i S H_3$ and $\Omega$ is the holomorphic 3-form of $X$. $W_0(U,S)$ depends on the dilaton through $G_3$ and it is also a function of the complex structure moduli since $\Omega$ depends on $U_\alpha$.

In general the scalar potential in supergravity is defined as:
\be
V = e^K \left[K^{a \bar{b}} D_a W D_{\bar{b}} \overline{W} - 3 |W|^2 \right],
\ee
where $a$ and $b$ run over all the moduli of the compactification and the K\"ahler covariant derivative is defined as $D_a W\equiv\partial_a W+ W \partial_a K$. The particular dependence of \eqref{leadingKP} on $\vo$ is such that the scalar potential obeys the \textit{no-scale structure}, namely:
\be
K^{i \bar{j}} \partial_i K \partial_{\bar{j}} K =  3 \,,
\ee
with $i$ and $j$ running over the K\"ahler moduli. Due to the no-scale structure and the fact that $W_0(U,S)$ does not depend on the $T$-moduli, the scalar potential simplifies to:
\be
V = e^K K^{\alpha \bar{\beta}} D_\alpha W D_{\bar{\beta}} \overline{W}\,,
\ee
where $\alpha$ and $\beta$ now run over the complex structure moduli and the dilaton. It is positive definite, so that $U_\alpha$ and $S$ can be stabilised supersymmetrically at:
\be
D_S W_0(U,S) = 0 \qquad\text{and}\qquad D_{U_{\alpha}} W_0(U,S) = 0\,,
\label{sustabilisation}
\ee
while the K\"ahler moduli remain unstabilised at this order of approximation.

\subsection{No-scale breaking effects}

In this section we will consider the complex structure and the dilaton stabilised at leading order as in \eqref{sustabilisation}, and we will introduce new effects to stabilise the K\"ahler moduli by breaking the no-scale structure. In particular we will consider two explicit setups:
\bi
\item[a)] \textit{Single modulus case} \\
In this setup we have a single K\"ahler modulus $T_b = \tau_b + {\rm i} c_b$ where $c_b$ is an axion field and $\tau_b$ controls the overall volume: $\vo = \tau_b^{3/2}$. The canonical normalisation of the inflaton field $\tau_b$ can be inferred from its kinetic terms: 
\be
\mc{L}_{\rm kin} = \frac{3}{4 \tau_b^2} \partial^\mu \tau_b \partial_\mu \tau_b\,,
\ee
so that the canonically normalised volume field can be written as:
\be
\Phi = \sqrt{\frac{3}{2}} \ln \tau_b \simeq \sqrt{\frac{2}{3}} \ln \vo\,.
\label{CanNorm}
\ee
In this setup the visible sector can be realised in two different ways:
\bi
\item[i)] The visible sector lives on a stack of D7-branes wrapping the 4-cycle associated to $\tau_b$. Given that the visible sector gauge coupling is set by $\tau_b$ as $\alpha_{\rm vis}^{-1}=\tau_b$, the late-time value of the volume has to be of order $100$ in order to reproduce a correct phenomenological value of $\alpha_{\rm vis}^{-1}\simeq 25$. This model is characterised by high scale SUSY.

\item[ii)] The visible sector is localised on D3-branes at singularities obtained by collapsing a blow-up mode to zero size due to D-term stabilisation \cite{Seq}. In this case the visible sector coupling is set by the dilaton, and so the volume can take much larger values of order $\vo \simeq 10^7$ which lead to TeV-scale supersymmetry via sequestering effects. In the presence of desequestering perturbative or non-perturbative effects \cite{Conlon:2010ji}, low-energy SUSY requires larger values of $\vo$ of order $\vo\sim 10^{14}$.
\ei

\item[b)] \textit{Two moduli case}\\
In this setup we start with two K\"ahler moduli: $T_b = \tau_b + {\rm i} c_b$ and $T_s = \tau_s + {\rm i} c_s$ with $\tau_b \gg \tau_s$. The Calabi-Yau volume takes the Swiss-Cheese form:
\be
\vo = \tau_b^{3/2} - \tau_s^{3/2}\,.
\ee 
The modulus which plays the role of the inflaton will still be $\tau_b$, and so its canonical normalisation is given by (\ref{CanNorm}). 
However, in this setup the visible sector can be realised in three different ways:
\bi
\item[i)] The visible sector lives on a stack of D7-branes wrapping $\tau_b$ (for an explicit Calabi-Yau construction see \cite{Cicoli:2011qg}). As explained above, in this case $\vo$ has to be of order $100$ for phenomenological reasons and the SUSY scale is very high.

\item[ii)] The visible sector lives on a stack of D7-branes wrapping $\tau_s$ (see again \cite{Cicoli:2011qg} for an explicit model). Given that the visible sector gauge coupling is now independent on $\tau_b$ since $\alpha_{\rm vis}^{-1}=\tau_s$, the volume can take large values. A particularly interesting value is $\vo\simeq 10^{14}$ which leads to TeV-scale soft terms, as in standard non-sequestered LVS models. 

\item[iii)] The visible sector is localised on D3-branes at singularities (see \cite{ExplSeq} for explicit global dS models) obtained by collapsing a blow-up mode to zero size due to D-term stabilisation. Since the gauge kinetic function of D3-branes is given by $S$, $\vo$ can take large values. Due to sequestering effects, in this case TeV-scale superpartners require $\vo\simeq 10^7$.
\ei
\ei

\subsubsection{$\alpha'$ corrections}

If $W_0\neq 0$, the no-scale structure is broken by $\alpha'^3$-corrections which show up in the K\"ahler potential in the following way \cite{Becker:2002nn}:
\be
K \supset - 2 \ln\left(\vo + \frac{\hat{\xi}}{2}\right)\simeq -2\ln\vo - \frac{\hat\xi}{\vo}\quad\Rightarrow\quad K_{\alpha'} = -\frac{\hat\xi}{\vo}\,,
\label{alphaprimeKP}
\ee
where $\hat{\xi} = \xi s^{3/2}$. Note that this correction causes a mixing between the K\"aler moduli and the dilaton which produces a small shift in the dilaton VEV of order $D_S W_0(U,S) \sim \mc{O}\left(\vo^{-1}\right)$.
The contribution of this $\alpha'$ correction to the scalar potential looks like: 
\be
V = e^K \left(K^{T_b \bar{T}_b} D_{T_b} W D_{\bar{T}_b} \overline{W} - 3 |W|^2 \right) \supset V_0 \,e^{-\sqrt{\frac{27}{2}} \Phi} \equiv V_{\alpha'}\,,
\label{Va}
\ee
where we have defined:
\be
V_0 = \frac{g_s}{8 \pi} \frac{3 \hat{\xi} |W_0|^2}{4}\,.
\label{eq:V0}
\ee
$V_0$ is a free parameter which can be tuned to get inflection point inflation with the right COBE normalisation and a large volume minimum.

\subsubsection{Non-perturbative effects}

Given that non-perturbative effects are exponentially suppressed, they tend to give rise to negligible contributions to the scalar potential. However, in the two moduli case, $T_s$-dependent non-perturbative effects could lead to potentially large contributions. In this case the superpotential becomes \cite{Kachru:2003aw}:
\be
W = W_0 + A_s(U,S)\, e^{- a_s T_s}\,,
\label{LVSsuperpot}
\ee
where $A_s(U,S)$ is an unknown function of the dilaton and the $U$-moduli. Given that these moduli have already been stabilised at leading order, $A_s$ can be considered as an $\mc{O}(1)$ constant. The parameter $a_s$ is given by $a_s = 2 \pi/N$, where $N$ is the number of D-branes wrapping the small cycle where the non-perturbative effect is generated.

The new effect in (\ref{LVSsuperpot}) breaks the no-scale structure by producing two terms in the scalar potential which allow to stabilise the K\"ahler moduli by competing with the $\alpha'^3$ effect (\ref{Va}). This stabilisation mechanism is at the basis of the so-called Large Volume Scenario \cite{Balasubramanian:2005zx}. Given that $\tau_s$ turns out to be much heavier than $\tau_b$, it can be integrated out in order to get effectively a scalar potential which depends only on $\tau_b$.

The LVS potential takes the form:
\be
V = \frac{g_s}{8 \pi} \left(\frac{8  a_s^2 A_s^2 \sqrt{\tau_s} e^{-2 a_s \tau_s}}{3 \V} - \frac{4 W_0 a_s A_s \tau_s e^{-a_s \tau_s}}{\V^2} + \frac{3 \hat{\xi} W_0^2}{4 \V^3}\right).
\label{eq:LVS}
\ee
Minimising with respect to $\tau_s$, we get at leading order in the $a_s \tau_s \sim \ln \V$ expansion:
\be
e^{-a_s \tau_s} \simeq \frac{3 W_0 \sqrt{\tau_s}}{4 a_s A_s \V}\,.
\ee 
Plugging this expression into (\ref{eq:LVS}) we get an effective potential which depends only on the volume $\vo$:
\be
V = \frac{V_0}{\vo^3} \left[1 - \frac{2}{\hat\xi a_s^{3/2}} \left(\ln \V\right)^{3/2}\right]\,.
\label{effectivepotentialLVS}
\ee
The first term in (\ref{effectivepotentialLVS}) is just the $\alpha'$ correction (\ref{Va}) whereas the second term leads to a new non-perturbative contribution to the scalar potential:
\be
V_{\rm np} = - \kappa_{\rm np} V_0 \,\Phi^{3/2}\,e^{-\sqrt{\frac{27}{2}} \Phi} \qquad\text{with}\qquad 
\kappa_{\rm np}=\frac{2}{\xi}\left(\frac{3}{8\pi^2}\right)^{3/4}\left(g_s N\right)^{3/2}\,.
\label{Vnp}
\ee
Note that $\kappa_{\rm np}$ is another parameter that can be tuned to get inflection point inflation with the right COBE normalisation and number of e-foldings.

\subsubsection{Higher derivative $\alpha'$ corrections}

Additional contributions to the scalar potential from higher derivative corrections to the 10D action have been computed in \cite{Ciupke:2015msa}. These corrections have the same higher dimensional origin as the $\alpha'^3$-term (\ref{Va}). Both stem from the $\alpha'^3 \mc{R}^4$ contribution to the in 10D type IIB action. The effect of these four derivative corrections is to modify the equations of motion of the auxiliary fields (F-terms) thereby giving rise to corrections to the kinetic terms, new quartic derivative couplings and more importantly new contributions to the scalar potential.  These $F^4$ corrections depend on the Calabi-Yau topology and take the generic form:
\be
V_{F^4} = - \frac{\hat{\lambda} |W_0|^4}{\V^4} \Pi_i t^i\,,
\label{v1}
\ee
where $\Pi_i$ are topological integers defined as:
\be
\Pi_i = \int_X c_2 \wedge \hat{D}_i\,,
\ee
with $c_2$ the second Chern class of the Calabi-Yau $X$ and $\hat{D}_i$ is a basis of harmonic $(1,1)$-forms allowing for the usual expansion of the K\"ahler form $J$ as:
\be
J = \sum_{i=1}^{h_{1,1}} t_i \hat{D}_i\,,
\ee
with $t^i$ being 2-cycle volumes. The parameter $\hat{\lambda}$ is expected to be of order $\hat{\xi}/\chi(X)$, with $\chi(X)$ being the Calabi-Yau Euler number. For model building purposes, we will take it to be a real negative constant.

Depending on the details of the compactification space, these new terms can take different forms, yielding contributions to the scalar potential that scale differently with the overall volume. In the simplest single K\"ahler modulus case where $t_b = \sqrt{\tau_b} \simeq \V^{1/3}$, the correction \eqref{v1} takes the form:
\be
V_{F^4} =  \frac{\kappa_{F^4}\,V_0}{\vo^{11/3}} = \kappa_{F^4}\, V_0\,e^{-\frac{11}{\sqrt{6}} \Phi}\,,
\label{VF4}
\ee
where we defined:
\be
\kappa_{F^4} \equiv - \hat{\lambda}  \Pi_b \frac{|W_0|^4}{V_0}\,.
\ee
The form of these corrections for the two moduli case can be derived by focusing on the $\mathbb{C} \rm{P}^4_{[1,1,1,6,9]}$ case where the volume can be written in terms of the 2-cycle volumes $t_1$ and $t_5$ as \cite{Denef:2004dm}:
\be
\vo = \frac{1}{6} (3 t_1^2 t_5 + 18 t_1 t_5^2 + 26 t_5^3)\,,
\ee
implying that the 4-cycles are given by:
\be
\tau_1 = \frac{\partial \V}{\partial t_1} = t_5 (t_1 + 3 t_5) \quad\text{and}\quad \tau_5 = \frac{\partial \V}{\partial t_5} = \frac{1}{2} (t_1 + 6 t_5)^2\,.
\ee
One can define $\tau_4$ as the linear combination:
\be
\tau_4 = \tau_5 - 6 \tau_1 = \frac{t_1^2}{2} \quad \Rightarrow \quad t_1 = \sqrt{2 \tau_4}\,.
\ee
Plugging $t_1$ back into the expression for $\tau_5$, solving the equation for $t_5$ and requiring that $t_5 > 0$ when $\tau_5 \gg \tau_4$ we get:
\be
t_5 = \frac{1}{3 \sqrt{2}} \left(\sqrt{\tau_5} - \sqrt{\tau_4}\right),
\ee
from which it can be inferred that the volume has the form:
\be
\vo = \frac{1}{9 \sqrt{2}} \left(\tau_b^{3/2} - \tau_s^{3/2}\right),
\ee
where we identified $\tau_5 \equiv \tau_b$ and $\tau_4 \equiv \tau_s$. Thus the $\alpha'$-correction of \eqref{v1} becomes:
\be
V_{F^4} \simeq V_0 \left(\frac{\kappa_{F^4_{(b)}}}{\vo^{11/3}} + \frac{\kappa_{F^4_{(s)}} \sqrt{\tau_s}}{\V^4}\right),
\label{deltav1}
\ee
where:
\be
\kappa_{F^4_{(b)}} = - \hat\lambda  \Pi_5 \frac{|W_0|^4}{6^{1/3} V_0} \qquad\text{and}\qquad \kappa_{F^4_{(s)}} = -\hat\lambda \sqrt{2} \left(\Pi_4 - \frac{\Pi_5}{6}\right) \frac{|W_0|^4}{V_0} \,.
\ee

\subsubsection{String loop corrections}

The tree-level K\"ahler potential \eqref{leadingKP}, in addition to $\alpha'$ corrections, can also receive corrections due to string loops. 
These are encoded into two contributions to the K\"ahler potential, accounting for the exchange of Kaluza-Klein and winding modes \cite{loops}:
\be
K^{KK}_{g_s} = g_s \sum_{i = 1}^{h_{1,1}} \frac{c_i (a_{ij} t^j)}{\V} \qquad\text{and}\qquad K^W_{g_s} = \sum_{i = 1}^{h_{1,1}} \frac{d_i (a_{ij} t^j)^{-1}}{\V}\,,
\ee
where $c_i$ and $d_i$ are unknown constants while $a_{ij}$ are combinatorial factors. The final contribution to the scalar potential can be written as:
\be
V_{g_s} = \sum_{i = 1}^{h_{1,1}} \frac{|W_0|^2}{\vo^2} \left(g_s^2 c_i^2 \frac{\partial^2 K_{\rm tree}}{\partial \tau_i^2} - 2 K^W_{g_s}\right).
\label{slccontributiontov}
\ee
Noting that $\partial^2 K_{\rm tree}/\partial\tau_b^2 = 3/(4 \tau_b^2)$, it turns out that in both scenarios the leading order string loop correction looks like:
\be
V_{g_s} = \frac{\kappa_{g_s}\,V_0}{\vo^{10/3}} = \kappa_{g_s}\,V_0\,e^{-\frac{10}{\sqrt{6}} \Phi}\,,
\label{Vgs}
\ee
where $\kappa_{g_s}$ is a real tunable number.

\subsubsection{Anti D3-branes}

Anti D3-branes yield a positive contribution to the scalar potential which in general provides a viable mechanism to realise a dS minimum. More precisely, the introduction of anti D3-branes in the compactification produces a term in the scalar potential of the form \cite{Kachru:2003aw}:
\be
V_{\overline{D3}} = \frac{\kappa_{\overline{D3}}\,V_0}{\vo^2} = \kappa_{\overline{D3}}\,V_0\,e^{-\sqrt{6}\Phi}\,,
\label{VD3}
\ee
where $\kappa_{\overline{D3}}$ is a positive real number which can be tuned to realise inflection point inflation with viable post-inflationary physics. 

\subsubsection{Charged hidden matter fields}

The possible presence on the big cycle of a hidden sector with matter fields $\phi$ charged under an anomalous $U(1)$ leads to the generation of moduli-dependent Fayet-Iliopoulos (FI) terms. The corresponding D-term potential reads:
\be
V_D = \frac{1}{2 \text{Re}(f_b)} \left(q_{\phi} |\phi|^2 - \xi_b\right)^2\,,
\ee 
where $f_b = T_b$ and $q_\phi$ is the $U(1)$-charge of $\phi$ while the FI-term is given by:
\be 
\xi_b = - \frac{q_b}{4 \pi} \frac{\partial K_{\rm tree}}{\partial T_b} = \frac{3 q_b}{8 \pi} \frac{1}{\vo^{2/3}}\,,
\ee 
where $q_b$ the $U(1)$-charge of $T_b$. Since supersymmetry breaking effects generate a mass for $\phi$ of order the gravitino mass, the total scalar potential becomes:
\be
V = V_D + c m_{3/2}^2 |\phi|^2 + \mc{O}\left(\vo^{-3}\right),
\ee
where $c$ is an $\mc{O}(1)$ coefficient which can be positive or negative depending on hidden sector model building details.
Integrating out $\phi$ leads to a new contribution which has been used to obtain dS vacua and takes the form \cite{ExplSeq}:
\be
V_{\rm hid} = \frac{\kappa_{\rm hid}\,V_0}{\vo^{8/3}} = \kappa_{\rm hid}\,V_0\,e^{-\frac{8}{\sqrt{6}}\Phi}\,,
\label{vds2}
\ee
where $\kappa_{\rm hid} = \frac{3 c q_b W_0^2}{16 \pi q_\phi V_0}$ is a tunable coefficient.\footnote{Note that $q_b$ and $q_\phi$ must have the same sign otherwise the minimum for $|\phi|$ would be at zero.}

\subsubsection{Total scalar potential}

The total scalar potential that we shall consider can in general be written as:
\be
V_{\rm tot} = V_{\alpha'} + V_{\rm np} +  V_{F^4} + V_{g_s} + V_{\overline{D3}} + V_{\rm hid}\,,
\label{Vtot}
\ee
where $V_{\alpha'}$ is the universal $\alpha'$ correction (\ref{Va}), $V_{\rm np}$ is the non-perturbative generated potential (\ref{Vnp}) which is non-negligible only in the two moduli case, $V_{F^4}$ are the higher derivative effects (\ref{VF4}) and (\ref{deltav1}), $V_{g_s}$ is the string loop potential (\ref{Vgs}), $V_{\overline{D3}}$ is the contribution (\ref{VD3}) from anti D3-branes and $V_{\rm hid}$ is the potential (\ref{vds2}) generated by the F-terms of charged hidden matter fields. 

Let us now add all these different contributions to the total scalar potential for the single modulus and the two moduli cases separately: 
\bi
\item[a)] \textit{Single modulus case}\\
In this simple model with only a single K\"ahler modulus, the generic expression for the scalar potential is:
\be
V(\Phi) = V_0 \left(e^{-\sqrt{\frac{27}{2}} \Phi} + \kappa_{g_s} \,e^{- \frac{10}{\sqrt{6}} \Phi} + \kappa_{F^4} \,e^{- \frac{11}{\sqrt{6}} \Phi} 
+ \kappa_{\overline{D3}} \,e^{- {\sqrt{6}} \Phi} + \kappa_{\rm hid} \,e^{- \frac{8}{\sqrt{6}} \Phi}\right).
\label{V1}
\ee
In this setup the post-inflationary dS minimum is generated by the interplay between the universal $\alpha'^3$ term and the two terms proportional to $\kappa_{\overline{D3}}$ and $\kappa_{\rm hid}$, with the inflationary inflection point arising at smaller volumes in a region where the first three terms in \eqref{V1} are comparable in size. 

\item[b)] \textit{Two moduli case}\\
In the two moduli case, the total scalar potential (\ref{Vtot}) contains at least six terms. However, as we shall see in the next section, we need just five tunable parameters in order to get inflection point inflation. We shall therefore neglect the last term in (\ref{Vtot}) which might be removed by a model building choice. We stress that this choice does not affect our final results. In fact, if instead we neglected $V_{\overline{D3}}$ in (\ref{Vtot}), we would obtain qualitatively the same results. Thus in this case the total inflationary potential becomes:
\be
V(\Phi) = V_0 \left[\left(1 - \kappa_{\rm np} \Phi^{3/2}\right) e^{-\sqrt{\frac{27}{2}} \Phi} + \kappa_{g_s} \,e^{-\frac{10}{\sqrt{6}} \Phi} 
+ \kappa_{F^4_{(b)}} \,e^{-\frac{11}{\sqrt{6}} \Phi} + \kappa_{\overline{D3}}\, e^{- \sqrt{6} \Phi}\right].
\label{V2}
\ee
Here we are including only the leading term of the higher derivatives corrections \eqref{deltav1} which is proportional to $\kappa_{F^4_{(b)}}$. Hence we are assuming that the term proportional to $\kappa_{F^4_{(s)}}$ is either very suppressed (a natural possibility given its volume scaling) or exactly vanishing.
\ei

\section{Single field dynamics}
\label{SecSingle}

In this section we study the effective single field inflationary dynamics for both the single modulus and the two moduli case. A deeper analysis of the effect of the heavy field for the two moduli case will be performed in Sec. \ref{SecTwo}. In order to obtain a phenomenologically viable model, we should require that:
\ben
\item There is an inflection point at $\Phi_{\rm ip}$.
\item The potential is such that the COBE normalisation is satisfied and the number of e-foldings is $N_e \simeq 60$.
\item There is de Sitter minimum at large volumes at $\Phi_{\rm min}$.
\een
As we will see below, these requirements translate into five conditions on the scalar potential, and so we need five tunable free parameters.

\subsection{Analytical discussion}

As a first step, let us discuss the strategy used to determine the free parameters in \eqref{V1} and \eqref{V2} in order to get inflection point inflation. The position of the inflection point $\Phi_{\rm ip}$ and the minimum $\Phi_{\rm min}$ can be chosen independently with the only constraint (apart from $\Phi_{\rm min}>\Phi_{\rm ip}$) being:
\be
\vo_{\rm ip} = e^{\sqrt{\frac{3}{2}} \Phi_{\rm ip}} \gtrsim 10^3\qquad\Leftrightarrow\qquad\Phi_{\rm ip} \gtrsim 5\,,
\ee
in order to trust the effective field theory during inflation.

Once $\Phi_{\rm ip}$ and $\Phi_{\rm min}$ are chosen, we impose that the scalar potential actually produces an inflection point and the late time minimum at the desired positions. This can be done by scanning over flux parameters, intersections numbers and gauge groups so that the following constraints are satisfied:
\bi
\item \textit{Inflection point}
\bea
&& (1) \quad \left.V''\right|_{\Phi = \Phi_{\rm ip}} = 0 \label{eq:cond1}\\
&& (2) \quad \left. V' \cdot V'''\right|_{\Phi = \Phi_{\rm ip}} = \left.\frac{2 \pi^2 V^2}{(170)^2}\right|_{\Phi = \Phi_{\rm ip}}\label{eq:cond2}
\eea
\item \textit{Late time minimum}
\bea 
&& (3) \quad \left.V'\right|_{\Phi = \Phi_{\rm min}} = 0 \label{eq:cond3}\\
&& (4) \quad \left.V\right|_{\Phi = \Phi_{\rm min}} = 0 \label{eq:cond4}
\eea
\ei
The first two conditions produce an inflection point at $\Phi_{\rm ip}$ with the right slope to yield a scalar spectral tilt around $n_s = 0.96$, while the last two conditions imply the existence of a Minkowski minimum at $\Phi_{\rm min}$. These conditions are invariant under a rescaling of $V_0$ in \eqref{V1} and \eqref{V2} since it is just an overall multiplicative factor in the scalar potential. $V_0$ is instead fixed by the requirement of obtaining the right COBE normalisation given by \eqref{cobe} which can also be rewritten as:
\be
(5) \quad \Delta^2 = \left.\frac{1}{24 \pi^2} \frac{V}{\epsilon}\right|_{\Phi = \Phi_*} \simeq \left.\frac{1}{12 \pi^2} \frac{V^3}{\left(V'\right)^2}\right|_{\Phi = \Phi_{\rm ip}} \simeq 2.4 \cdot 10^{-9}\,.
\ee
Given the definition of $V_0$ in \eqref{eq:V0}, condition (5) can be seen as a constraint on the magnitude of the flux superpotential $W_0$ which in type IIB string compactifications naturally lies in the range $[0.1,\,100]$. We will show now how additional constraints on $\Phi_{\rm ip}$ arise from combining this naturalness criterion with the requirement of low-energy supersymmetry. In fact we shall carefully choose $\Phi_{\rm ip}$ so that the COBE normalisation fixes $W_0$ in the natural range mentioned above. Fixing $W_0$ through the condition (5) sets also the energy scale of the soft terms. Here we distinguish between two possibilities:
\bi
\item \textit{Non-sequestered models}: If the cycle supporting the visible sector is stabilised in geometric regime, the soft terms are of order the gravitino mass:
\be
M_{\rm soft} \simeq m_{3/2} \simeq \sqrt{\frac{g_s}{8 \pi} }\frac{W_0}{\vo_{\rm min}}\,.
\label{eq:unseq}
\ee
This is usually referred to as the non-sequestered case which for $W_0\sim 1$ leads to TeV-scale supersymmetry only for values of the volume as large as $\vo\sim 10^{14}$. This is possible for the two moduli case where the value of the visible sector coupling does not depend on $\vo$. In the single modulus case, since $\alpha_{\rm vis}^{-1}=\vo^{2/3}$, the volume has to be of order $100$, resulting necessarily in a high-scale SUSY scenario.

\item \textit{Sequestered models}: If the visible sector modulus is fixed in the singular regime, the soft terms can be very suppressed with respect to the gravitino mass: 
\be
M_{\rm soft} \simeq \frac{m_{3/2}}{\vo} \simeq \sqrt{\frac{g_s}{8 \pi}} \frac{W_0}{\vo_{\rm min}^2}\,.
\label{eq:seq}
\ee
This is usually referred to as the sequestered scenario which for $W_0\sim 1$ leads to low-energy supersymmetry only for $\vo\sim 10^7$. In these sequestered models supersymmetry is broken in the bulk of the extra-dimensions while the visible sector lives on branes localised at a singularity. Given that in this case the visible sector coupling is set by the dilaton, this scenario can be realised both in the single and in the two moduli case.
\ei
The magnitude of the flux superpotential that satisfies condition (5) for a generic inflection point $\Phi_{\rm ip}$ can be estimated by noting that at horizon exit $\epsilon\sim 10^{-10}$ and there is a percent level cancellation between the three dominant terms. This implies that at the inflection point $V\sim 0.01\ V_0 \ e^{-\sqrt{27/2}\Phi_{\rm ip}}$, and so the Hubble scale can be estimated as:
\be
\label{hubbleestimate}
H = \sqrt{\frac{V_{\rm ip}}{3 M_P^2}} \simeq \frac{1}{10 \sqrt{3}} \frac{\sqrt{V_0 M_P^2}}{\vo_{\rm ip}^{3/2}} = \sqrt{\frac{\xi}{2 \pi g_s^{1/2}}} \frac{W_0 M_P}{40 \vo_{\rm ip}^{3/2}}\,,
\ee
where we used \eqref{eq:V0} and we restored the correct dependence on the Planck mass $M_P$. Using the same argument we can also estimate the amplitude of the scalar perturbations in \eqref{cobe} as:
\be
\left.\frac{1}{24 \pi^2} \frac{V}{\epsilon}\right|_{\Phi = \Phi_*}\simeq \frac{1}{24 \pi^2} \frac{0.01\ V_0 \ e^{-\sqrt{27/2}\Phi_{\rm ip}}}{10^{-10}}
\simeq 2.4 \times 10^{-9}\,,
\ee
which yields:
\be
W_0^2\simeq 7.58 \times 10^{-15}\ 8 \pi \sqrt{g_s} \, \, e^{\sqrt{27/2}\ \Phi_{\rm ip}}\,.
\label{eq:w0}
\ee
Assuming $g_s\simeq 0.1$, we conclude that only inflection points in the range $\Phi_{\rm ip} \in [6,\,10]$ are compatible with natural values of $W_0$.

We can take these simple estimates further and for each $\Phi_{\rm ip}$ obeying \eqref{eq:w0} find the position of the late time minimum $\Phi_{\rm min}$ that gives rise to TeV-scale soft masses in both sequestered and non-sequestered scenarios. For the sequestered case we have:
\be
\frac{g_s}{8 \pi}\ W_0^2 \ e^{-2 \sqrt{6}\ \Phi_{\rm min}}\simeq 10^{-30}\,,
\ee
which by using \eqref{eq:w0} becomes:
\be
\Phi_{\rm min}\simeq 6.76 +\frac{3}{4}\Phi_{\rm ip}\,.
\label{eq:est_seq}
\ee
A similar estimate for the non-sequestered case yields:
\be
\Phi_{\rm min}\simeq 13.51 +\frac{3}{2}\Phi_{\rm ip}\,.
\label{eq:est_unseq}
\ee
Hence we see that the distance between the inflection point and the minimum in the non-sequestered case is exactly twice the corresponding distance in the sequestered setup. The factor of two descends directly from the extra volume suppression of \eqref{eq:seq} when compared with \eqref{eq:unseq}. In both cases the combination of the observational constraint on the amplitude of the density perturbations, the theoretical bias on natural values of $W_0$ and the requirement of TeV-scale soft terms conspire to fix the distance between the inflationary inflection point and the late-time minimum.

Tables \ref{tab1} and \ref{tab2} illustrate some reference values obtained using \eqref{eq:w0} to fix the inflection point for different values of $W_0$, and then \eqref{eq:est_unseq} and \eqref{eq:est_seq} to get the late-time minimum in the non-sequestered (Tab. \ref{tab1}) and sequestered (Tab. \ref{tab2}) scenarios respectively.

\begin{table}[h!]
\begin{center}
\begin{tabular}{cccccc}
\hline
$W_0$ & $\Phi_{\rm ip}$ & $\vo_{\rm ip}$ & $\Phi_{\rm min}$ & $\vo_{\rm min}$  \\
\hline
$0.1$ & $7.03$ & $5.5 \times 10^3$ & $24.06$  & $6.3 \times 10^{12}$ \\
\hline
$1$ & $8.29$ & $2.5 \times 10^4$ & $25.94$ & $6.3 \times 10^{13}$ \\
\hline
$10$ & $9.54$ & $1.2 \times 10^5$ & $27.82$ & $6.3 \times 10^{14}$ \\
\hline
$100$ & $10.79$ & $5.5 \times 10^5$ & $29.70$ & $6.3 \times 10^{15}$ \\
\hline
\end{tabular}
\end{center}
\caption{Positions of the inflection point and the late-time minimum for the non-sequestered case obtained by requiring a correct COBE normalisation and low-energy supersymmetry for natural values of $W_0$ and setting $g_s=0.1$.}
\label{tab1}
\end{table}

\begin{table}[h!]
\begin{center}
\begin{tabular}{cccccc}
\hline
$W_0$ & $\Phi_{\rm ip}$ & $\vo_{\rm ip}$ & $\Phi_{\rm min}$ & $\vo_{\rm min}$  \\
\hline
$0.1$ & $7.03$ & $5.5 \times 10^3$ & $12.03$  & $2.5 \times 10^{6}$ \\
\hline
$1$ & $8.29$ & $2.5 \times 10^4$ & $12.97$ & $7.9 \times 10^{6}$ \\
\hline
$10$ & $9.54$ & $1.2 \times 10^5$ & $13.91$ & $2.5 \times 10^{7}$ \\
\hline
$100$ & $10.79$ & $5.5 \times 10^5$ & $14.85$ & $7.9 \times 10^{7}$ \\
\hline
\end{tabular}
\end{center}
\caption{Positions of the inflection point and the late-time minimum for the sequestered case obtained by requiring a correct COBE normalisation and low-energy supersymmetry for natural values of $W_0$ and setting $g_s=0.1$.}
\label{tab2}
\end{table}

Using the values listed in Tab. \ref{tab1} and \ref{tab2} to estimate the Hubble scale as in \eqref{hubbleestimate} we get $H \simeq 10^{10}$ GeV which corresponds to an inflationary scale of order $10^{14}$ GeV. This result can also be obtained numerically using the more precise values listed in the next section.

Let us comment on the consistency of our effective field theory approach. As derived in \cite{Cicoli:2013swa}, the superspace derivative
expansion is under control if $m_{3/2}\ll M_{\rm KK}$ which translates into the bound:
\be
\delta \equiv \sqrt{\frac{g_s}{2}} \frac{W_0}{\vo^{1/3} } \ll 1\,.
\ee
This bound is satisfied in each case of Tab. \ref{tab1} and \ref{tab2} both around the inflection point and the late-time minimum. In fact, considering just the region around the inflection point ($\vo$ becomes larger around the minimum and so this bound is stronger during inflation) and setting $g_s=0.1$, we have: $\delta\simeq 10^{-3}$ for $W_0=0.1$, $\delta\simeq 10^{-2}$ for $W_0=1$, $\delta\simeq5\cdot 10^{-2}$ for $W_0=10$ and $\delta\simeq 0.1$ for $W_0=100$.
Thus the superspace derivative expansion is under control, and so higher derivative $\alpha'$ corrections should naturally be suppressed. Therefore, as we shall show in the next section, we have to tune the coefficient of the $F^4$ $\alpha'$ terms (\ref{VF4}) and (\ref{deltav1}) to large values. However the fact that the expansion parameter $\delta$ turns out to be small, allows us to neglect further higher derivative corrections in a consistent way.

\subsection{Numerical results}

\begin{figure}[H]
\centering
\includegraphics[width=.6\textwidth]{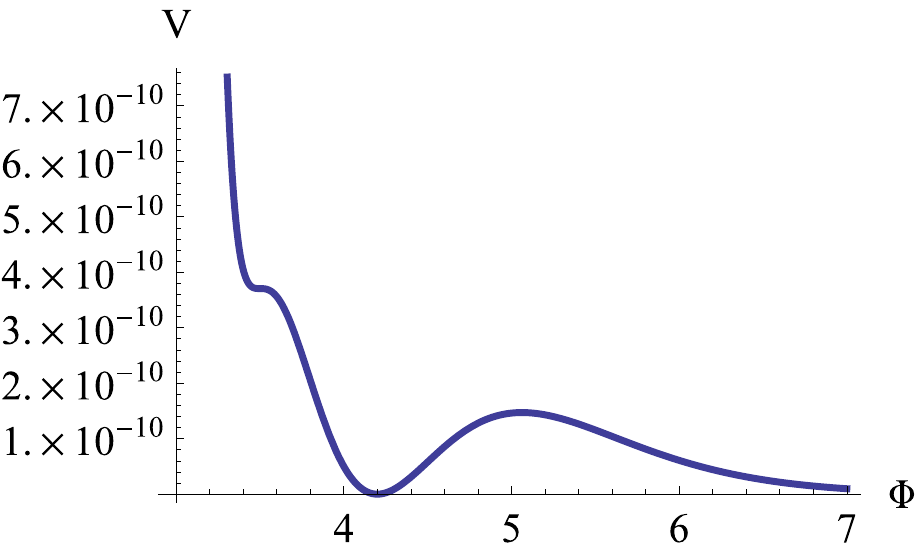}
\caption{Scalar potential obtained requiring $\Phi_{\rm ip} = 3.5$ and $\Phi_{\rm min} = 4.2$ in the single modulus case with the visible sector on D7-branes wrapping the volume cycle.}
\label{smallvol}
\end{figure}

In this section we present a detailed numerical study of inflection point volume inflation. We start by focusing on the non-sequestered single modulus case with high-scale SUSY. In this case the late-time minimum is bound to be of order $100$, so that the evolution of the canonically normalised field $\Phi$ is very limited. Nevertheless, as shown in Fig. \ref{smallvol}, it is possible to get an inflection point and a late-time minimum at the desired values. The scalar potential is plotted in Fig. \ref{smallvol}. We require $\Phi_{\rm ip} = 3.5$ and $\Phi_{\rm min} = 4.2$, corresponding respectively to values of the volume $\vo_{\rm ip} \simeq 72$ and $\vo_{\rm min} \simeq 171$. These values are clearly too small to trust the effective field theory approach. However we shall still present the numerical results for this case for illustrative purposes, and shall focus later on cases with low-energy supersymmetry where the volume during and after inflation is larger and the supergravity effective theory is under much better control.

In this example it is possible to reproduce the correct value of the spectral index, as a consequence of the condition \eqref{eq:cond2}. We list in Tab. \ref{tab0} the numerical results obtained for different positions of the inflection point and the late-time minimum for which we get always $n_s \simeq 0.96$.
\begin{table}[H]
\begin{center}
\begin{tabular}{cccccccc}
\hline
$\Phi_{\rm ip}$ & $\Phi_{\rm min}$ & $W_0$ & $\kappa_{g_s}$ & $\kappa_{F^4}$ & $\kappa_{\overline{D3}}$ & $\kappa_{\rm hid}$ & $\Delta \Phi/M_{\rm P}$ \\
\hline
$2.5$ & $3.8$ & $3 \times 10^{-5}$ & $-2.55$ & $2.26$ & $1.05 \times 10^{-3}$ & $-0.14$  & $0.21$\\
\hline
$3$ & $4$ & $8 \times 10^{-5}$ & $-2.98$ & $3.11$  & $7.09 \times 10^{-4}$ & $-0.12$ & $0.17$ \\
\hline
$3.5$ & $4.2$ & $2 \times 10^{-4}$ & $-3.48$ & $4.28$ & $4.71 \times 10^{-4}$ & $-0.10$ & $0.12$ \\
\hline
\end{tabular}
\end{center}
\caption{Numerical results for the coefficients of the scalar potential for the single modulus case with the visible sector on D7-branes wrapping the volume cycle.}
\label{tab0}
\end{table}

$\Delta \Phi$ is the field excursion of the canonically normalised volume modulus $\Phi$ between horizon exit and the end of inflation. Since $\Delta \Phi\sim 0.1 M_P$ we are clearly dealing with a small field inflationary model. Thus the tensor-to-scalar ratio is of order $r \simeq 10^{-10}$. The values of $W_0$ reported in Tab. \ref{tab0} are the numerical results which satisfy the COBE normalisation. The corresponding Hubble scale in each case is $H \simeq 10^9$ GeV which translates into an inflationary scale around $10^{14}$ GeV. 

\begin{figure}
\centering
\subfloat[][\emph{Inflection point}.]
{\includegraphics[width=.45\textwidth]{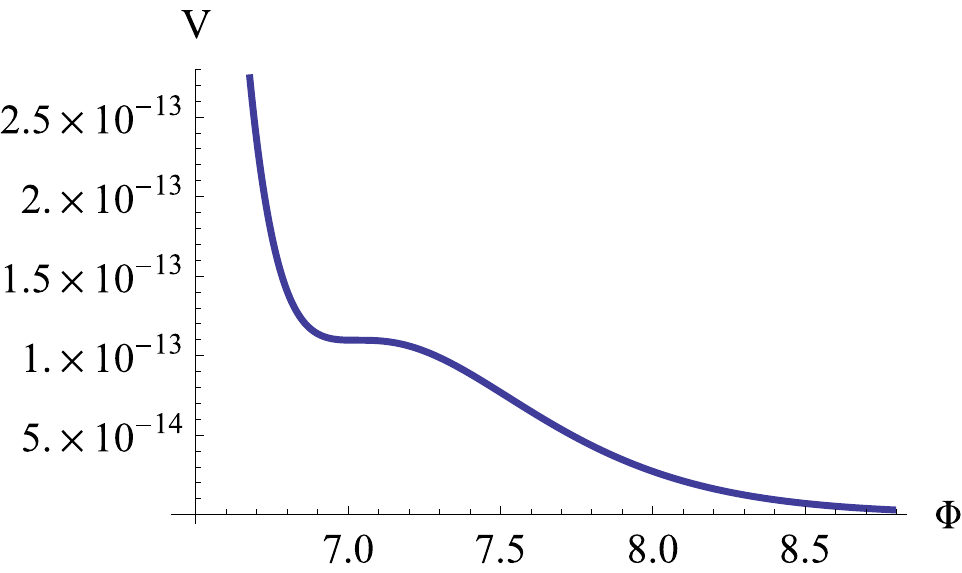}} \quad
\subfloat[][\emph{Late time minimum}.]
{\includegraphics[width=.45\textwidth]{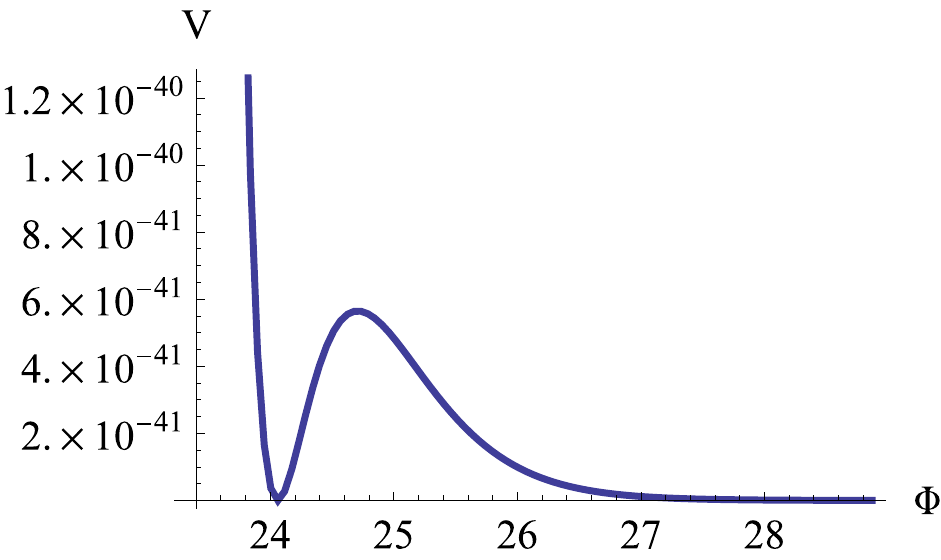}} 
\caption{Scalar potential for non-sequestered models with TeV-scale supersymmetry obtained requiring $\Phi_{\rm ip} = 7.03$ and $\Phi_{\rm min} = 24.06$.}
\label{SingleFieldNonSequestered1}
\end{figure}

We now turn to study the two more interesting sequestered and non-sequestered cases with larger values of the volume and TeV-scale supersymmetry. 
Since in both cases the shape of the scalar potential is always qualitatively the same around the inflection point and the late-time minimum, we plot it in Fig. \ref{SingleFieldNonSequestered1} just for the non-sequestered case. 

The numerical results listed below are obtained by requiring that the inflection point and the late-time minimum are those given in Tab. \ref{tab1} and \ref{tab2} where we required natural values of $W_0$ and low-energy supersymmetry in both non-sequestered and sequestered cases. The tensor-to-scalar ratio turns out to be always of order $r \simeq 10^{-9}$. In the following tables the only inputs are $\Phi_{\rm ip}$ and $\Phi_{\rm min}$, while $W_0$, $\kappa_{g_s}$, $\kappa_{F^4}$ (or $\kappa_{F^4_{\rm (b)}}$), $\kappa_{\rm hid}$, $\kappa_{\overline{D3}}$ and $\kappa_{\rm np}$ are the numerical outputs obtained by solving \eqref{eq:cond1} and \eqref{eq:cond4}. As a consequence of the condition \eqref{eq:cond2}, the value of the spectral index is $n_s \simeq 0.967$ in each of the cases listed below.

\bi
\item \textit{Single modulus case}: 

Let us start with the simplest single modulus case of \eqref{V1}. The numerical results relative to the non-sequestered case with low-energy SUSY are listed in Tab. \ref{tab3} while those relative to the sequestered case are listed in Tab. \ref{tab4}:
\begin{table}[H]
\begin{center}
\begin{tabular}{cccccccc}
\hline
$\Phi_{\rm ip}$ & $\Phi_{\rm min}$ & $W_0$ & $\kappa_{g_s}$ & $\kappa_{F^4}$ & $\kappa_{\overline{D3}}$ & $\kappa_{\rm hid}$ & $\Delta \Phi/M_{\rm P}$ \\
\hline
$7.03$ & $24.06$ & $0.06$ & $-31.68$ & $253.85$ & $7.94 \times 10^{-14}$ & $-8.11 \times 10^{-5}$  & $0.42$\\
\hline
$8.29$ & $25.94$ & $0.6$ & $-53.02$ & $710.59$ & $7.93 \times 10^{-15}$ & $-3.76 \times 10^{-5}$ & $0.42$ \\
\hline
$9.54$ & $27.82$ & $6.2$ & $-88.35$ & $1972.74$ & $7.94 \times 10^{-16}$ & $-1.74 \times 10^{-5}$ & $0.42$ \\
\hline
$10.79$ & $29.70$ & $62.1$ & $-147.21$ & $5476.08$ & $7.94 \times 10^{-17}$ & $-8.12 \times 10^{-6}$ & $0.42$ \\
\hline
\end{tabular}
\end{center}
\caption{Numerical results for the coefficients of the scalar potential for the non-sequestered single modulus case with TeV-scale supersymmetry.}
\label{tab3}
\end{table}

\begin{table}[H]
\begin{center}
\begin{tabular}{cccccccc}
\hline
$\Phi_{\rm ip}$ & $\Phi_{\rm min}$ & $W_0$ & $\kappa_{g_s}$ & $\kappa_{F^4}$ & $\kappa_{\overline{D3}}$ & $\kappa_{\rm hid}$ & $\Delta \Phi/M_{\rm P}$ \\
\hline
$7.03$ & $12.03$ & $0.07$ & $-25.46$ & $187.38$ & $1.30 \times 10^{-7}$ & $-8.47 \times 10^{-3}$ & $0.39$ \\
\hline
$8.29$ & $12.97$ & $0.79$ & $-41.50$ & $504.98$ & $3.92 \times 10^{-8}$ & $-5.59 \times 10^{-3}$ & $0.38$ \\
\hline
$9.54$ & $13.91$ & $8.14$ & $-67.27$ & $1346.28$ & $1.17 \times 10^{-8}$ & $-3.68 \times 10^{-3}$ & $0.38$ \\
\hline
$10.79$ & $14.85$ & $83.6$ & $-108.80$ & $3575.02$ & $3.48 \times10^{-9}$ & $-2.42 \times 10^{-3}$ & $0.37$ \\
\hline
\end{tabular}
\end{center}
\caption{Numerical results for the coefficients of the scalar potential for the sequestered single modulus case with TeV-scale supersymmetry.}
\label{tab4}
\end{table}

\item \textit{Two moduli case}: 

Now we turn to the two moduli setup of \eqref{V2}. The results for the non-sequestered case are listed in Tab. \ref{tab5} while the results for the sequestered case are presented in Tab. \ref{tab6}.
\begin{table}[H]
\begin{center}
\begin{tabular}{cccccccc}
\hline
$\Phi_{\rm ip}$ & $\Phi_{\rm min}$ & $W_0$ & $\kappa_{g_s}$ & $\kappa_{F^4_{\rm (b)}}$ & $\kappa_{\overline{D3}}$ & $\kappa_{\rm np}$ & $\Delta \Phi/M_{\rm P}$ \\
\hline
$7.03$ & $24.06$ & $0.07$ & $-25.42$ & $193.01$ & $8.46 \times 10^{-15}$  & $8.91 \times 10^{-3}$ & $0.41$\\
\hline
$8.29$ & $25.94$ & $0.75$ & $-41.34$ & $525.04$ & $7.83 \times 10^{-16}$ & $7.93 \times 10^{-3}$ & $0.41$ \\
\hline
$9.54$ & $27.82$ & $7.55$ & $-67.16$ & $1421.51$ & $7.29 \times 10^{-17}$ & $7.12 \times 10^{-3}$ & $0.41$ \\
\hline
$10.79$ & $29.70$ & $75.8$ & $-109.33$ & $3858.01$ & $6.82 \times 10^{-18}$ & $6.43 \times 10^{-3}$ & $0.41$ \\
\hline
\end{tabular}
\end{center}
\caption{Numerical results for the coefficients of the scalar potential for the non-sequestered two moduli case with TeV-scale supersymmetry.}
\label{tab5}
\end{table}

\begin{table}[H]
\begin{center}
\begin{tabular}{cccccccc}
\hline
$\Phi_{\rm ip}$ & $\Phi_{\rm min}$ & $W_0$ & $\kappa_{g_s}$ & $\kappa_{F^4_{\rm (b)}}$ & $\kappa_{\overline{D3}}$ & $\kappa_{\rm np}$ & $\Delta \Phi/M_{\rm P}$ \\
\hline
$7.03$ & $12.03$ & $0.12$ & $-15.55$ & $97.00$ & $2.48 \times 10^{-8}$  & $2.28 \times 10^{-2}$ & $0.38$ \\
\hline
$8.29$ & $12.97$ & $1.27$ & $-23.17$ & $236.06$ & $6.86 \times 10^{-9}$ & $2.02 \times 10^{-3}$ & $0.37$ \\
\hline
$9.54$ & $13.91$ & $13.7$ & $-34.53$ & $572.48$ & $1.89 \times 10^{-9}$ & $1.80 \times 10^{-2}$ & $0.37$ \\
\hline
$10.79$ & $14.85$ & $147.9$ & $-51.58$ & $1390.83$ & $5.25 \times 10^{-10}$ & $1.62 \times 10^{-2}$ & $0.36$ \\
\hline
\end{tabular}
\end{center}
\caption{Numerical results for the coefficients of the scalar potential in the sequestered two moduli case with TeV-scale supersymmetry.}
\label{tab6}
\end{table}

Two important observations can be inferred from the values of the coefficients listed in Tab. \ref{tab3}, \ref{tab4}, \ref{tab5} and \ref{tab6}. The first one is that $\kappa_{g_s}$ and $\kappa_{\rm hid}$ are always required to be negative. In our models the negative sign of $\kappa_{g_s}$ can be obtained by the interplay of the two terms in \eqref{slccontributiontov} while the sign of $\kappa_{\rm hid}$ depends on hidden sector model building details. Moreover the presence of the inflection point is highly sensitive to small variations of the coefficients in Tab. \ref{tab3}, \ref{tab4}, \ref{tab5} and \ref{tab6}. Thus in order to accurately reproduce the shape of the scalar potential in each case, it is necessary to tune the coefficients to a much higher level of precision than that reported in the Tables.
\ei

\section{Two field dynamics}
\label{SecTwo}

Up to this point we have dealt exclusively with the single field limit, implicitly assuming that all other moduli, like the axio-dilaton, the complex structure moduli and additional K\"ahler moduli, are heavier than the Hubble scale during inflation. While this may be arranged for in the single K\"ahler modulus case, it is certainly not true for the model constructed within the LVS framework. In this section we comment on various aspects of the two field dynamics of this model.

\begin{figure}[h!]
\centering
\subfloat[][\emph{Inflationary region}.]
{\includegraphics[width=.47\textwidth]{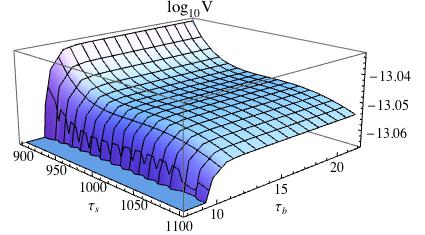}} \quad
\subfloat[][\emph{Large volume region}.]
{\includegraphics[width=.47\textwidth]{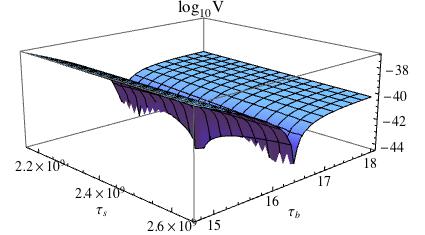}} 
\caption{Scalar potential \eqref{V2} in the inflationary region, around the tuned point $(\tau_b,\tau_s)|_{\rm ip}=(1000,20)$ and in the large volume region $(\tau_b,\tau_s)|_{\rm min}=(2.4\times 10^9, 16.4)$.}
\label{fig:2FieldLVS_pot}
\end{figure}

As shown in Sec. \ref{SecOrigin}, the effective single field potential of \eqref{V2} is obtained after integrating out the small blow-up modulus $\tau_s$. This procedure is valid in the vicinity of the LVS minimum where there is a clear mass hierarchy:
\be
m_{\tau_s}^2 \sim \frac{g_s}{8\pi}\frac{W_0^2}{\vo_{\rm min}^2} \gg \frac{g_s}{8\pi}\frac{W_0^2}{\vo_{\rm min}^3} \sim m_{\tau_b}^2\,,
\ee
however it fails around the inflection point where both fields are very light $m_{\tau_s}^2, m_{\tau_b}^2 \ll H^2 $, implying that both will be dynamical during inflation. This is illustrated in Fig. \ref{fig:2FieldLVS_pot}.

Due to the flatness of the scalar potential along both directions, the correct way to analyse the system is by numerically solving the background field equations as done in the original work \cite{Conlon:2008cj}. Here we will extend the aforementioned analysis by studying the sensitivity to the choice of initial conditions in a given potential and by clarifying to what extent the single field results constitute a valid approximation to the inflationary observables. The reader looking for more details of the setup is referred to \cite{Conlon:2008cj} as we will focus only on the results obtained.

We proceed in the same spirit of the single field analysis by choosing the coefficients that induce an inflection point along the volume direction. For concreteness we choose $(\tau_b,\tau_s)|_{\rm ip}=(1000,20)$. We then consider a set of initial conditions around that point and numerically solve the equations of motion. In Fig. \ref{fig:2FieldLVS} we plot the solutions for the different choices of initial positions for the system. We assume throughout that the fields are released with vanishing velocities and find that the sensitivity to the initial position, that is characteristic of inflection point models, is magnified in the two field setup as perturbing the initial conditions by a small amount can lead to drastically different outcomes. Starting uphill from the inflection point tends to lead to trajectories that produce insufficient expansion. In some cases, depending on the ratio $\tau_s^{\rm ip}/\tau_s^{\rm min}$, some of these trajectories can lead to the collapse of the compact manifold to vanishing volume. For trajectories that start downhill from the tuned point, it is easier to obtain a viable post inflationary evolution, with the system evolving towards the LVS minimum, but the number of e-foldings decreases drastically with the distance from the tuned point. One is therefore led to the conclusion that viable inflationary trajectories are obtained only in a narrow region around $\tau_b^{\rm ip}$ and $\tau_s\gtrsim \tau_s^{\rm ip}$.

\begin{figure}[h!]
\centering
{\includegraphics[width=.47\textwidth]{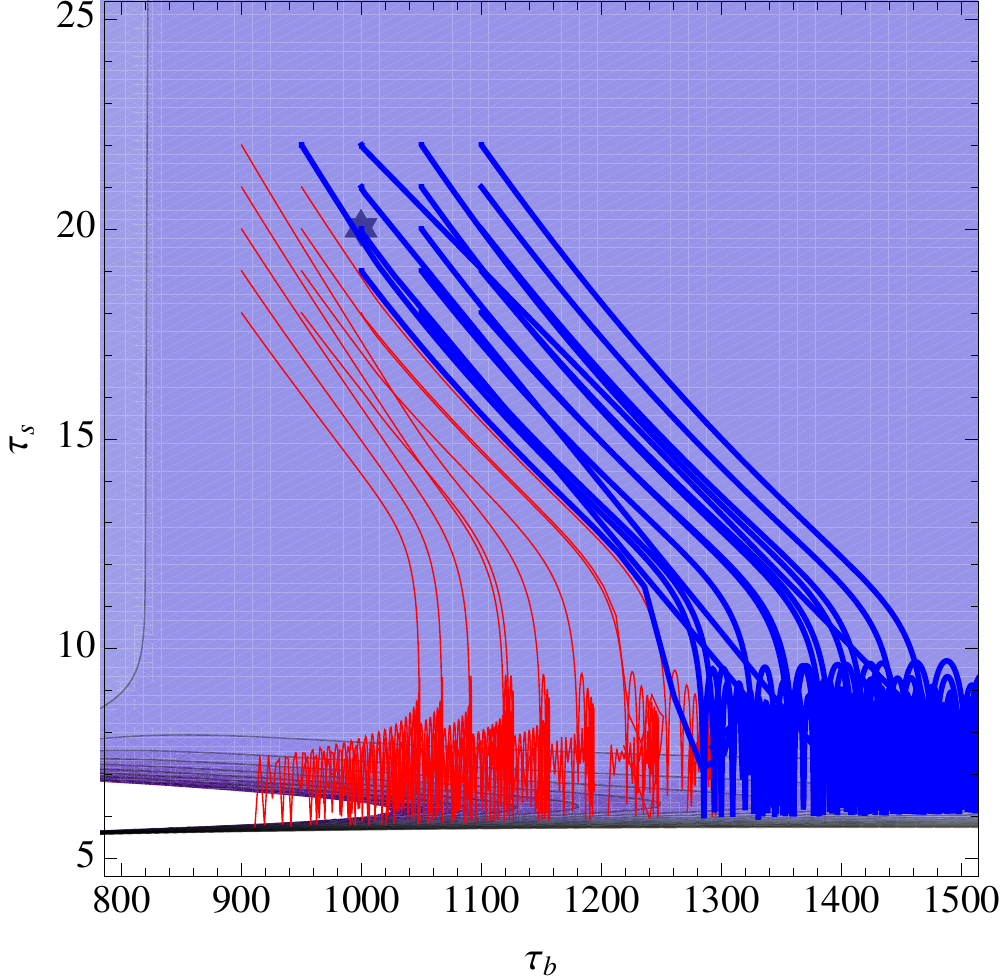}} \quad
{\includegraphics[width=.47\textwidth]{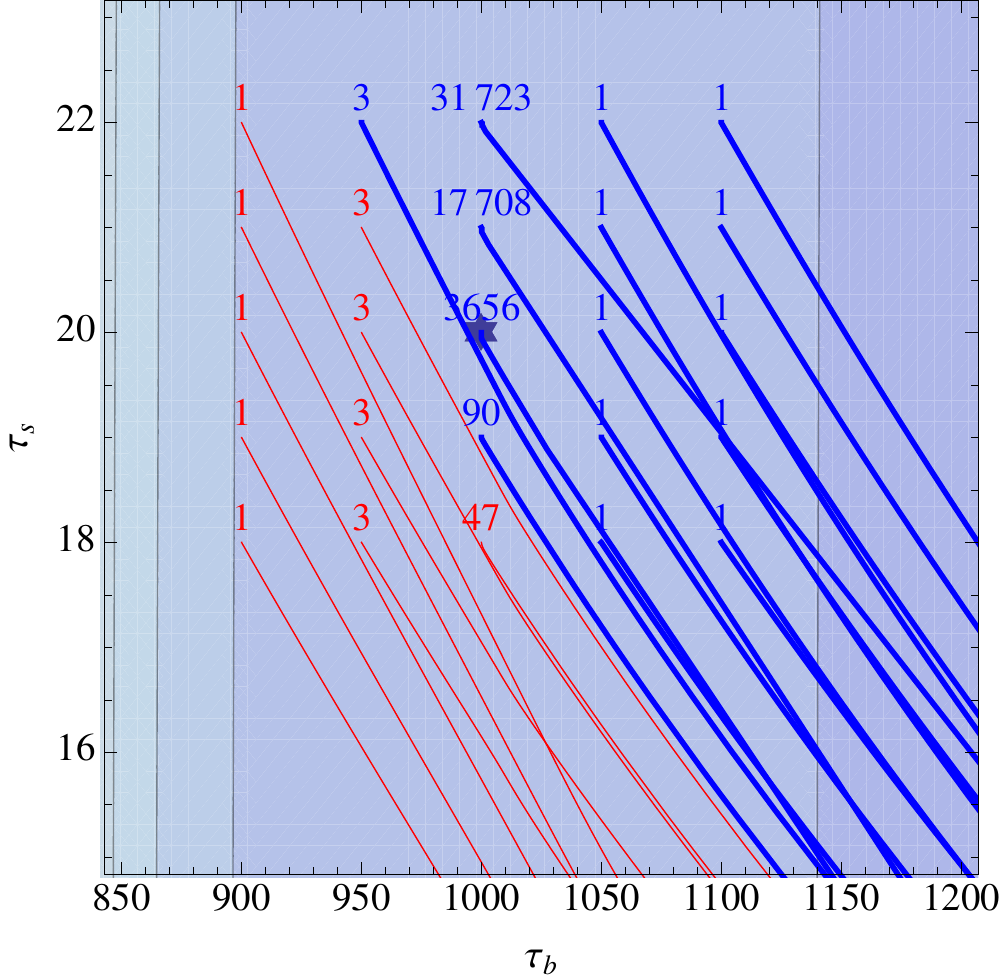}} 
\caption{Trajectories around the flat inflection point at $(\tau_b,\tau_s)= (1000 , 20)$ marked by the star. Blue (thicker) trajectories correspond to those evolving towards the post-inflationary large volume minimum, while the magenta (thinner) end with a collapsing volume modulus. Right: Magnification around the initial points. The numbers denote the total number of slow-roll e-foldings for each trajectory.}
\label{fig:2FieldLVS}
\end{figure}

In what concerns inflationary observables in the two field setup one expects the projection along the inflationary trajectory (and hence the single field estimates of Sec. \ref{SecSingle}) to be a good approximation to the full result. This follows directly from the fact that the observable portion of the trajectories yielding $N_e\ge 60$, is rather straight, as can be seen by the smallness of the inverse curvature radius plotted in Fig. \ref{fig:2FieldLVS}. This implies that curvature and isocurvature perturbations are essentially decoupled and that a straightforward generalisation of the single field case leads to an accurate estimate of the cosmological observables.\footnote{For thorough discussion of this issue in the context of a local string inflation model see \cite{Bielleman:2015lka}.}
\begin{figure}[h!]
\centering
{\includegraphics[width=.5\textwidth]{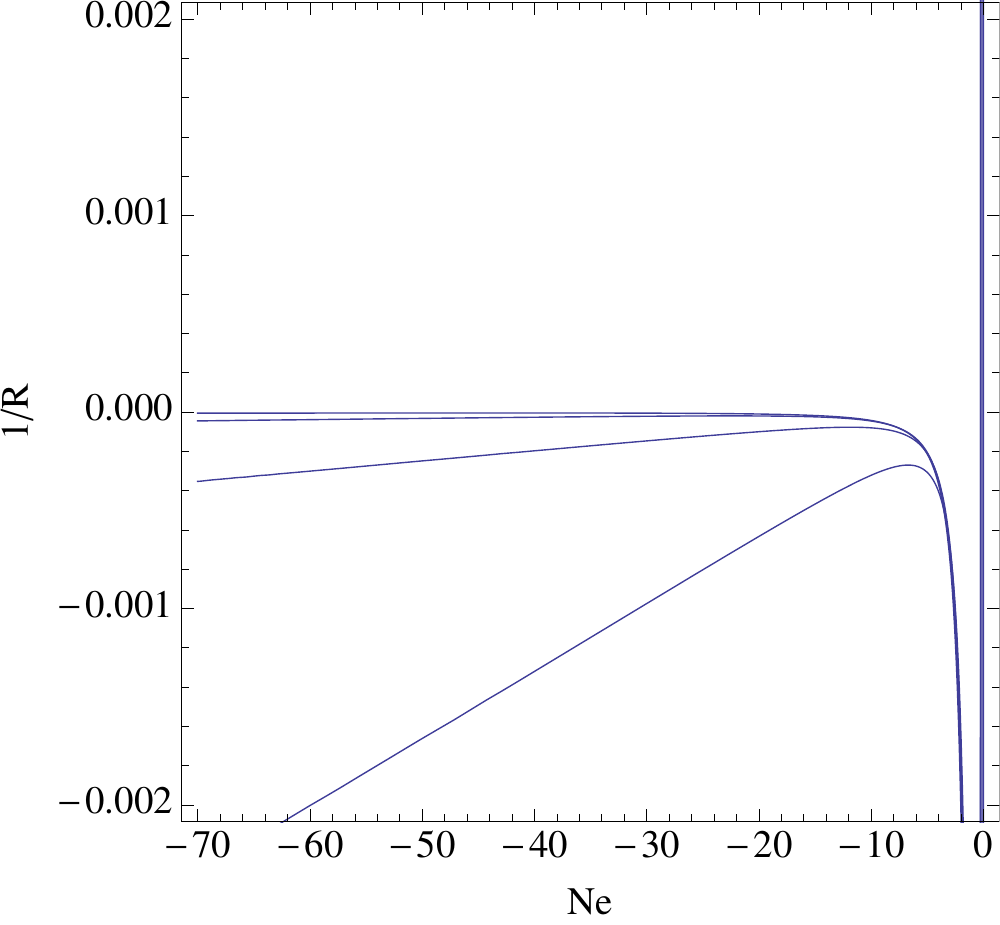}} 
\caption{Inverse curvature radius for trajectories leading to $N_e>60.$}
\label{fig:2FieldLVS}
\end{figure}

\section{Conclusions}
\label{SecConcl}

It has long been acknowledged that there is severe tension between having simultaneously low-scale supersymmetry and high scale inflation in supergravity models. This is due to the fact that the inflationary energy density contributes to the moduli potential and will in generally tend to destabilise them.

Several mechanisms to decrease or eliminate this tension have been proposed over the years and in this work we further develop the proposal of \cite{Conlon:2008cj}. This particular model solves the tension between TeV-scale supersymmetry and the inflationary scale by having an evolving compactification volume between the inflationary epoch and today. Inflation is due to an inflection point in the volume direction of the scalar potential which also possesses a minimum at large volume where the modulus is supposed to sit at late-time after inflation. In order to prevent overshooting it is necessary to require that a small amount of radiation is generated after the end of inflation. The presence of this additional radiation is well justified in the two field model since it could be produced by the oscillations of the heavy modulus around its minimum while in the single field model particle production could be induced by a changing vacuum state between the end of inflation and today. 

In this paper we described a possible microscopic origin of the inflationary scalar potential that allows the gravitino mass to vary after the end of inflation and at the same time features a late-time dS minimum. In particular, we provided an explicit construction where the inflationary inflection point is generated by the interplay between string loops and higher derivative $\alpha'$ corrections to the scalar potential. Moreover, we supplemented the LVS construction with a new model that involves only one K\"ahler modulus. While in the LVS two moduli model non-perturbative effects play a crucial r\^ole in determining the presence of both the inflection point and the late-time minimum, in the single modulus model non-perturbative effects are absent and an additional contribution arising from the F-terms of charged hidden matter is needed. For both models we analysed the relation between the value of the volume during inflation and at present with the size of the flux superpotential $W_0$. We found that the distance between the inflection point and the late-time minimum is fixed by the choice of $W_0$. Requiring a late-time minimum which leads to TeV-scale supersymmetry, natural values of the flux superpotential $W_0\sim \mc{O}(1)$ give rise to inflection point inflation around values of the volume of order $\vo\sim\mc{O}(10^5)$ which is large enough to trust the effective field theory approach. 

We finally studied the full dynamics of the two field system in the LVS model and showed that, after tuning the potential such that it features the desired inflection point, there is a significant sensitivity to the choice of initial conditions. Perturbing the starting positions of the fields even by a small amount can lead to a radically different cosmological evolution. We also showed that, despite the presence of two dynamical fields, the predictions for the cosmological observables derived in the single field case are accurate since the field space trajectories are essentially straight over the last $60$ e-foldings of expansion.

\section*{Acknowledgements}

FM would like to thank the Instituto de Fisica Teorica (IFT UAM-CSIC) in Madrid for its support via the Centro de Excelencia Severo
Ochoa Programme under Grant SEV-2012-0249. This work has been supported by the ERC Advanced Grant SPLE under contract ERC-2012-ADG-20120216-320421, by the grant FPA2012-32828 from the MINECO, and the grant SEV-2012-0249 of the \textit{Centro de Excelencia Severo Ochoa} Programme.

\end{document}